\documentclass[a4paper,11pt]{article}

\usepackage{color}
\usepackage{ulem}
\usepackage{hyperref}
\usepackage{enumerate}
\usepackage{latexsym}
\usepackage{eucal}
\usepackage{appendix}

\def \be {\begin{equation}}
\def \ee {\end{equation}}
\def \ba {\begin{array}}
\def \ea {\end{array}}
\def \bea {\begin{eqnarray}}
\def \eea {\end{eqnarray}}

\def \arccosh {{\rm arccosh}}
\def \arctanh {{\rm arctanh}}

\def \psiright {{|\Psi_\alpha\rangle_\Lambda}}
\def \psileft  {{~_\Lambda \langle \Psi_\alpha|}}

\def \psiradialright{{|\Psi_\alpha(\rho_0,0,t)\rangle}}
\def \psiradialleft {{\langle \Psi_\alpha(\rho_0,0,t)|}} 
\def \mo   {{\mathcal{O}_\alpha}}
\def \psilocalright {{|\psi_\alpha(z_0,\bar z_0)\rangle_\epsilon}}
\def \psilocalleft {{~_\epsilon\langle \psi_\alpha(z_0,\bar z_0)|}}
\def \hmp {{\hat \mathcal{P}_\rho}}
\def \hms {{\hat \mathcal{S}_\phi}}
\def \hmt {{\hat \mathcal{T}_\phi}}
\def \hmq {{\hat \mathcal{Q}_\rho}}
\begin{document}

\title{Position and momentum operators for a moving particle in bulk}
\author{Wu-zhong Guo\footnote{wuzhong@hust.edu.cn}}
\date{}


\maketitle

\vspace{-10mm}
\begin{center}
{\it School of Physics, Huazhong University of Science and Technology\\
Luoyu Road 1037, Wuhan, Hubei 430074, China\vspace{1mm}
}
\vspace{10mm}
\end{center}
\begin{abstract}
In this paper we explore how to describe a bulk moving particle in the dual conformal field theories (CFTs). One aspect of this problem is to construct the dual state of the moving particle. On the other hand one should find the corresponding operators associated with the particle.  The dynamics of the particle, i.e., the geodesic equation, can be formulated  as a Hamiltonian system with canonical variables. The achievements of our paper are to construct the dual CFT states and the operators  corresponding to the canonical variables. The expectation values of the operators give the expected solutions of the geodesic line, and the quantum commutators reduce to the classical Poisson brackets to leading order in the bulk gravitational coupling. Our work provides a framework to understand the geodesic equation, that is gravitational attraction, in the dual CFTs.
\end{abstract}


\section{Introduction}
To understand gravity one should not only know the curved spacetime but also how matter moves in the spacetime. The AdS/CFT correspondence provides us a framework to explore both aspects of gravity in the conformal field theory (CFT) on the boundary of the asymptotically AdS spacetime\cite{Maldacena:1997re}-\cite{Witten:1998qj}. 

In the context of AdS/CFT, previous studies mainly focus on the first aspect, that is emergence of spacetime from the non-gravitational degrees of freedom.  The concepts from the quantum information theories are found to be useful. Many quantities, such as entanglement entropy, complexity, are expected to be related to the bulk geometry\cite{Ryu:2006bv}-\cite{Nguyen:2017yqw}.
One could refer to the recent review \cite{Bousso:2022ntt} for more references on these studies.

It is also significant to make clear how to describe the bulk moving particle in the CFTs. To answer this question, one should construct the dual state of the moving particle. Besides that, it is also necessary to construct the corresponding operators that are associated with the particle's position and momentum.

In this paper our motivation is to construct the dual state and the corresponding operators in the vacuum AdS$_3$ spacetime. There are no systematical methods to achieve the constructions. What we have done is to guess the right states and associated operators which could produce the correct classical geodesic line of the bulk particle at the leading order of $G$. Of course, we also have other checks. Our achievements are to construct a self-consistent description of the bulk moving particle in the CFT.

The constructions should follow some general rules:
\begin{enumerate}
\item The background geometry $g_{\mu\nu}$ can be effectively described by a CFT state $|g\rangle$. 
\item A bulk moving particle can be seen as excited state of the bulk, denoted by $|\psi\rangle$, which  is also a state  in the Hilbert space of the dual CFT. The energy of the particle should be given by $\langle\psi| \hat H_{cft}|\psi\rangle -\langle g|\hat H_{cft}|g\rangle $, where $\hat H_{cft}$ is the Hamiltonian operator of the CFT. 
\item There exists Hermitian operators corresponding to the canonical variables of the classical particle. The expectation values of these operators in the state $|\psi\rangle$ satisfy the equation of motion of the particle at the leading order of $G$. 
\item The quantum commutators of the constructed operators should reduce to the  classical Poisson brackets  in the semiclassical limit $G\to$ 0.
\end{enumerate} 

In the paper we only focus on the vacuum state of AdS$_3$. With the observation on the geodesic solution, we assume that the position and momentum operators can be constructed by stress energy tensor $T$ and $\bar T$, or equally the Virasoro generators. Actually, we will show below only global Virasoro generators are needed. 

We expect the operators would be state-dependent\cite{Papadodimas:2015jra}.
In the Hamiltonian formulation the canonical variables would be associated with the metric $g_{\mu\nu}$. According to the point 3 of the general rules, we would like to construct the operators associated with the canonical variables. It is nature to consider the operators should depend on the background geometry. However, they should not depend on the dual state of the particle, which is related to the initial conditions of the bulk particle. 

The organization of this paper is as follows. In section.\ref{geometric} we will briefly discuss the so-called geometric state. One of the  feature is that the correlators of stress energy tensor will satisfy the factorization property, which is important for our constructions. In section.\ref{geodesiclinesection}, the solution of the geodesic line in the global coordinate is shown. In section.\ref{Radialsection} the radial moving particle is discussed. The dual CFT state is assumed to be associated with the local bulk state with suitable regularization. We show how to construct the radial position and momentum operator in this case.  In section.\ref{locallysection} we discuss another example. The boundary locally excited state can be taken as particle starting from the AdS boundary. We also construct a new state that are expected to be dual to particle with angular momentum. The angular momentum of the particle has a dictionary with the rapidity of a boost in the CFT. The final section is conclusion and discussion. We discuss three interesting problems that are worthy to explore in the near future. 

\section{Geometric state and factorization }\label{geometric}
In the introduction we have mentioned that the background geometry is expected to be dual to a CFT state $|g\rangle$.  We will call these kinds of CFT states geometric states. The states $|\Psi\rangle$ that are expected to be dual to the moving particle should also be geometric states.  It is expected that the particle should have backreaction on the background geometry. If the backreaction can be neglected in the semiclassical limit $G\to 0$, it is not expected the observables in the CFT could detect the difference from the reference state $|g\rangle$. For example, the energy is same in  both states in the limit $G\to 0$. The construction would be meaningless. Therefore, in the following examples the mass of the particle will be taken to be $O(1/G)$.

We construct the position and momentum operators by using stress energy tensor $T$ and $\bar T$. One of the feature of the geometric states is the factorization property. For a given geometric state $|g\rangle$,  the expectation value of $T$ is of order $c$ or $1/G$\cite{Brown:1986nw}, for 2-point correlator
\bea
\langle g| T(z_1) T(z_2)|g\rangle-\langle g| T(z_1)|g\rangle \langle g| T(z_2)|g\rangle\sim O(c). 
\eea
Or we can define the scaled operator $u:=T/c$, the expectation value of which is of order $c^0$. The above condition becomes
\bea
\lim_{c\to \infty}\left(\langle g| u(z_1) u(z_2)|g\rangle-\langle g| u(z_1)|g\rangle \langle g| u(z_2)|g\rangle\right)=0.
\eea
This means the operators $u$ satisfy the factorization property in the limit $c\to \infty$. Actually, for n-point correlator we also have the factorization property. In \cite{Guo:2018fnv} we have shown the factorization condition is associated with the geometric state by the scaling behavior of holographic R\'enyi entropy. 

If the operators are functions of stress energy tensor, they also satisfy the factorization property. For an arbitrary operator $\hat X$ as a function of $T$ and $\bar T$, if the expectation value of it is finite in the semiclassical limit $c\to \infty$, we will call it classical operator.  Two arbitrary classical operators $\hat X$ and $\hat Y$, we expect the factorization
\bea\label{factorizationgeneral}
\lim_{c\to \infty}\left(\langle g | \hat X  \hat Y |g\rangle- \langle g | \hat X |g\rangle \langle g| \hat Y |g\rangle\right)=0.
\eea
For given geometric states and classical operators, such as the ones that we construct in the following, one could check the above statement by direct calculation. Actually, the factorization property is general for quantum system which has a well defined classical limit \cite{Yaffe:1981vf}.

The factorization property is very useful for our calculations.  For example, consider $\langle \hat X  \rangle_g:=\langle g| \hat X |g\rangle,\langle \hat Y\rangle_g:=\langle g| \hat Y |g\rangle \sim O(c^0)$. By the factorization  (\ref{factorizationgeneral}) we have $\langle \hat X\hat Y \rangle_g\sim O(c^0)$. Hence, we expect the commutator
\bea
\langle [\hat X,\hat Y ]\rangle_g \sim O(c^{-1}).
\eea
Now the commutator 
\bea
&&\langle[\hat X^2, \hat Y]\rangle_g=\langle \left(\hat X [\hat X, \hat Y]+[\hat X, \hat Y]\hat X\right) \rangle_g\nonumber \\
&&\phantom{\langle[\hat X^2, \hat Y]\rangle_g}=2 \langle\hat X [\hat X, \hat Y] \rangle_g+O(c^{-2})
\eea
More generally, one could check
\bea
\langle [\hat X^n, \hat Y]\rangle_g =n \langle \hat X^{n-1}[\hat X,\hat Y]\rangle_{g}+O(c^{-2}).
\eea
The above results will be used in the following sections.

\section{Geodesic line}\label{geodesiclinesection}
We will focus on the global coordinate. The metric is  
\bea\label{global}
ds^2=-\cosh^2(\rho)dt^2+d\rho^2+\sinh^2 (\rho)d\phi^2,
\eea
where we take the radius of AdS to be $1$.

Consider a particle starting from $(\rho_0,\phi_0)$ with velocity $\frac{d\phi}{dt}|_{t=0}=v_{\phi}$ and $\frac{d\rho}{dt}|_{t=0}=0$. The action of the  particle with mass $m$ is
\bea
S=\int dt L(t),
\eea 
where the Lagrangian is 
\bea
 L(t):=-m \sqrt{\cosh^2(\rho)-\dot{\rho}^2-\sinh^2 (\rho)\dot{\phi}^2}.
\eea
The canonical momentum associated with the coordinates $\rho$ and $\phi$ is
\bea\label{radialcanonical}
&&P_\rho:=\frac{\partial L}{\partial \dot{\rho}}=\frac{m\dot{\rho}}{\sqrt{\cosh^2(\rho)-\dot{\rho}^2-\sinh^2 (\rho)\dot{\phi}^2}}, \nonumber \\
&&P_\phi:= \frac{\partial L}{\partial \dot{\phi}}=\frac{m\sinh^2(\rho)\dot{\phi}}{\sqrt{\cosh^2(\rho)-\dot{\rho}^2-\sinh^2 (\rho)\dot{\phi}^2}}.
\eea
Using these we obtain the Hamiltonian 
\bea\label{globalenergy}
H=P_\rho \dot{\rho}+P_\phi \dot{\phi}-L=\frac{m\cosh^2(\rho)}{\sqrt{\cosh^2(\rho)-\dot{\rho}^2-\sinh^2 (\rho)\dot{\phi}^2}}.
\eea
The geodesic line of the particle can be obtained by solving the Hamiltonian equations associated with the canonical variables $\{ \rho,\phi,P_\rho,P_\phi\}$.
The result is 
\bea\label{radialsolution}
&&\tanh \rho(t)=\tanh (\rho_0) \sqrt{\left(1-v_\phi^2\right) \cos ^2(t)+v_\phi^2} \nonumber\\
&&\tan\phi(t)=v_\phi \tan(t)+\phi_0.
\eea
We can get the momentum $P_\rho(t)$ and $P_\phi(t)$  by taking the solutions (\ref{radialsolution}) into (\ref{radialcanonical})
\bea\label{canonicalmomentumglobal}
&&P_\rho(t)=\frac{m\left(1-v_{\phi }^2\right)\sinh (\rho_0) \sin (t) }{\sqrt{1+v_{\phi }^2\tan ^2(t) } \sqrt{1-v_{\phi }^2\tanh ^2(\rho_0) }},\nonumber \\
&&P_\phi(t)=\frac{mv_{\phi }\sinh \left(\rho _0\right) \tanh \left(\rho _0\right) }{\sqrt{1-v_{\phi }^2\tanh ^2\left(\rho _0\right)} }.
\eea
The Hamiltonian of the particle is conserved. Thus, the energy of the particle is constant, that is given by  
\bea\label{energyglobal}
E=\frac{m\cosh (\rho_0)}{\sqrt{1-v_\phi^2 \tanh^2 (\rho_0)}}.
\eea
Another constant of motion is the angular momentum $P_\phi$, which is independent with $t$ as we can see from (\ref{canonicalmomentumglobal}).

\section{Radial moving particle}\label{Radialsection}

Firstly, let us consider the radial moving particle, that is the velocity $v_\phi=0$. We would like to show  the dual CFT state of the radial moving particle. Then we will construct the position and momentum operators corresponding to the canonical variables $\{\rho,P_\rho\}$. 
\subsection{State dual to radial moving particle}

The bulk local states have been explored in many literatures. The Hamilton-Kabat-Lifschytz-Lowe (HKLL) construction is a well known method to express the bulk local operator as CFT operators \cite{Hamilton:2005ju}-\cite{Hamilton:2006fh}.  A different view on the construction is proposed in \cite{Miyaji:2015fia}, for which the symmetry of AdS and CFT play an important role\cite{Nakayama:2015mva}. We will briefly review the methods and show the bulk local states with suitable regularization can be taken as the dual state of the radial moving particle.

The bulk scalar operator $\hat{\phi}_\alpha(X^\mu)$ satisfies the equation of motion on the background geometry $g_{\mu\nu}(X^\mu)$,
\bea
(\Box^2_{g_{\mu\nu}}+m^2)\hat{\phi}_\alpha(X^\mu)=0,
\eea
where $m$ is the mass of scalar field.  Suppose the metric $g_{\mu\nu}$ can be associated with a geometric state $|\Psi(g_{\mu\nu})\rangle$. 
The bulk local state is defined as  $|\phi_\alpha(X^\mu)\rangle=\hat{\phi}_\alpha (X^\mu)|\Psi(g_{\mu\nu})\rangle$, where $X^\mu$ is the coordinate of the local operator. It is expected the bulk operator $\hat{\phi}_\alpha(X^\mu)$ can be expanded by the CFT operators. Thus the bulk local state $|\phi_\alpha(X^\mu)\rangle$ can be taken as states in Hilbert space of the CFT. 

We only focus on the vacuum state $|0\rangle$. Consider the global coordinate,
 the state located in the origin of AdS $\rho=0$, denoted by $|\Psi_\alpha\rangle$, can be expanded as the superposition of  Ishibashi states \cite{Miyaji:2015fia}
\bea
|\Psi_\alpha\rangle=\sum_{k=0}^\infty (-1)^{k}\frac{\Gamma(\Delta_\alpha)}{\Gamma(k+1)\Gamma(\Delta_\alpha+k)}L^{k}_{-1}\bar{L}^k_{-1}|\mathcal{O}_\alpha\rangle,
\eea
where $\Delta_\alpha=h_\alpha+\bar h_{\alpha}$ is the conformal dimension of primary operator $\mathcal{O}_\alpha$, the primary state $|\mathcal{O}_\alpha\rangle:=  \lim_{z\to 0}\mathcal{O}_\alpha |0\rangle$.  The standard AdS/CFT dictionary gives the relation $m= \sqrt{\Delta_\alpha (\Delta_\alpha -2)}$. 

The bulk local states at point $(\rho,\phi)$ can be associated with $|\Psi_\alpha\rangle$ by a unitary transformation $g(\rho,\phi)$.  The bulk local state at  point $(\rho,\phi)$  is given by 
\bea\label{gtransform}
|\Psi_\alpha(\rho,\phi)\rangle=g(\rho,\phi)|\Psi_\alpha\rangle,
\eea
with the unitary operator 
\bea\label{grang}
g(\rho,\phi)=e^{i(L_0-\bar L_0)\phi} e^{-\frac{\rho}{2}\left(L_1-L_{-1} +\bar L_1-\bar L_{-1}\right)}.
\eea

The state $|\Psi_\alpha\rangle$ is unnormalized since the local operator $\hat \phi_\alpha$ is unbounded operator. We can introduce a regulator $\Lambda$ and define the state
\bea
|\Psi_\alpha\rangle_\Lambda:= \mathcal{N}(\Lambda) e^{-\Lambda \hat H} |\Psi_\alpha\rangle,
\eea
where $\hat H:=L_0+\bar L_0$, the normalization constant
$\mathcal{N}(\Lambda)=e^{-\Lambda \Delta_\alpha}\sqrt{1-e^{-4\Lambda}}$. 
It is straightforward to obtain the following one-point functions of $L_n$
\bea\label{L0global}
\psileft L_0 \psiright=h_\alpha+\frac{1}{e^{4\Lambda}-1}
\eea
and $\psileft L_n \psiright =0$  for $n\neq 0$. It is also useful to evaluate the two-point functions
\bea
&&\psileft L_0^2 \psiright=h_{\phi }^2+\frac{2 h_{\phi }}{e^{4 \Lambda }-1}+\frac{e^{4 \Lambda }+1}{\left(e^{4 \Lambda }-1\right)^2},\nonumber \\
&&\psileft L_1L_{-1}\psiright=\frac{2 e^{4 \Lambda } h_{\phi }}{e^{4 \Lambda }-1}+\frac{2 e^{4 \Lambda }}{\left(e^{4 \Lambda }-1\right)^2},\nonumber \\
&& \psileft L_1^2\psiright=\psileft L_{-1}^2\psiright=0.
\eea
More generally, we have
\bea
\psileft L_0^m \psiright=h_\alpha^m+\sum_{k=1}^{m}h_\alpha^{m-k}C_{m}^k \sum_{n=0}^\infty n^{k} e^{-4\Lambda n}(1-e^{-4\Lambda}).
\eea
As we have argued in the beginning of section.\ref{geometric} we are interested in the case $\Delta_\alpha\sim O(c)$ in the holographic CFTs with $c\gg 1$. Define the operator $l_n:= L_n/c$, which can be taken as the classical operator. In the regime of $\Lambda \gg 1$, we would have the following clustering property for $l_n$,
\bea
\psileft l_n^m \psiright =\psileft l_n \psiright^m+O(c^{-1}), 
\eea
for $n=-1,0,+1$. This is a necessary condition for the geometric states as we have discussed in the introduction. Here we would like to explain the state $\psiright$ with $\Delta_\alpha \sim O(c)$ and $\Lambda \gg 1$ to be dual to a particle with mass $m$ at rest in the center of AdS in the global coordinate (\ref{global}). We have the  parameter relation $\Delta_\alpha\simeq m$.

The energy in the state $\psiright$ is given by $E=\psileft H_g \psiright=\Delta_\alpha-\frac{c}{12}+O(c^0)$, where $H_g:= L_0+\bar L_0 -\frac{c}{12}$ is the Hamiltonian of the boundary CFT in the global coordinate. The constant $-\frac{c}{12}$ is the Casimir energy in the vacuum of global coordinate.  This is consistent with the holographic result of a stationary particle with mass $m$ at $\rho=0$ by using the fact $\Delta_\alpha \simeq m$ at leading order of $c$. Actually, it is expected an object at rest in AdS$_3$ is dual to primary state $|\mathcal{O}_\alpha\rangle$ \cite{Kaplan}. In the limit $\Lambda \to \infty$, $\psiright$ would approach to $|\mathcal{O}_\alpha\rangle$. However, even taking $\Lambda \sim O(c^0)$ we find the bulk metric $\psiright$ still corresponds to the backreacted geometry with the stationary massive particle at $\rho=0$. The bulk metric is not sensitive to the cut-off parameter $\Lambda$. For our purpose we will take the state $\psiright$ with $\Lambda \ge O(c^0)$ as the stationary massive particle at $\rho=0$.

By using (\ref{gtransform}) the state of a particle located at $\rho=\rho_0,\phi=\phi_0$ is given by 
\bea
|\Psi_\alpha(\rho_0,\phi_0)\rangle=g(\rho_0,\phi_0)\psiright,
\eea
where $g(\rho,\phi)$ is given by (\ref{grang}). In the radial moving case one could always to fix the angular coordinate $\phi_0=0$.  Consider its time evolution we have the state
\bea\label{radialparticle}
|\Psi_\alpha(\rho_0,0,t)\rangle:=U_g(t)|\Psi_\alpha(\rho_0,0)\rangle,
\eea 
where $U_g(t):=e^{it H_g}$ is the unitary evolution operator.  In the following we would like to show this state is dual to a radial moving particle in the bulk by directly constructing the associated position and momentum operators.

\subsection{Position and Momentum operator}

 Let's calculate the expectation value of the Hamiltonian $H_g$ in the (\ref{radialparticle}). It is obvious that the energy $\psiradialleft H_g \psiradialright$ is independent with $t$.  To evaluate it we need the formula
\bea
g^{-1}(\rho_0,0)\  L_0 \ g(\rho_0,0)=L_0 \cosh(\rho_0)+\frac{L_1+L_{-1}}{2}\sinh(\rho_0).
\eea
By using (\ref{L0global}) we have
\bea\label{exphamradial}
\psiradialleft H_g \psiradialright=\Delta_\alpha \cosh (\rho_0) -\frac{c}{12}.
\eea
The first term is same with the classical particle energy (\ref{energyglobal}) with $v_\phi=0$ by taking $\Delta_\alpha \simeq m$. The second term is the Casimir energy in the vacuum.  The expectation value of the operator $\hat H= H_g+\frac{c}{12}$ gives energy of the particle. This suggests $\hat H$ can be taken as the operator dual to the Hamiltonian (\ref{globalenergy}).

One could also check the expectation value of the momentum operator $P_\phi:=L_0-\bar L_{0}$  is zero by using the fact $h_\alpha=\bar h_\alpha$. This is consistent with the result that $P_\phi=0$ for $v_\phi=0$.  

Now we move on to the construction of position operator $\hat{\rho}_r$ and momentum operator $\hmp$ of the radial moving particle. To simplify the notations we will denote the expectation value $\psiradialleft \hat X\psiradialright $   as $\langle \hat X\rangle_{\Psi_\alpha(t)}$ .

The basic requirement for the CFT operators $\hat \rho$ and $\hmp$ is that
\bea\label{basicrequirement}
&&\langle \hat \rho_r \rangle_{\Psi_\alpha(t)} = \rho(t),\nonumber \\
&& \langle \hmp \rangle_{\Psi_\alpha(t)} = P_\rho(t),
\eea
where $\rho(t)$ and $\hat P_\rho(t)$ are given by (\ref{radialcanonical}) with $v_\phi=0$. As we have discussed above the Hamiltonian $\hat H$ and momentum operator $P_\phi$ can be associated with energy and  angular momentum of the bulk moving particle. They are constructed by the Virasoro generators $L_n$ and $\bar L_{n}$. Motivated by this we can try to build $\hat{\rho}_r$ and $\hat{P}_\rho$ by the same way. Actually, we only need the generators associated with global conformal symmetry, that is $\{ L_{-1},L_{0},L_{1}\}$ and  $\{ \bar L_{-1},\bar L_{0},\bar L_{1}\}$. 

Firstly, let's show the following formulas that are useful for the constructions,
\bea\label{transformationlawradial}
&&g^{-1}(\rho_0,0)U_g^{-1}(t)  L_0 U_g(t) g(\rho_0,0)=g^{-1}(\rho_0,0) L_0  g(\rho_0,0),\nonumber \\
&&g^{-1}(\rho_0,0)U_g^{-1}(t)  L_1 U_g(t) g(\rho_0,0)\nonumber \\
&&=L_0\sinh(\rho_0)e^{it}+\frac{\cosh(\rho_0)e^{it}}{2}(L_1+L_{-1})+\frac{L_1-L_{-1}}{2}e^{it},\nonumber \\
&&g^{-1}(\rho_0,0)U_g^{-1}(t)  L_{-1} U_g(t)g(\rho_0,0)\nonumber \\
&&=L_0\sinh(\rho_0)e^{-it}+\frac{\cosh(\rho_0)e^{-it}}{2}(L_1+L_{-1})-\frac{L_1-L_{-1}}{2}e^{-it}.\nonumber \\
~
\eea 
It can be shown that
\bea
&&\langle\frac{L_1-L_{-1}}{2i} \rangle_{\Psi_\alpha(t)}=\psileft L_0 \psiright \sinh(\rho_0)\sin(t)\nonumber \\ 
&&\phantom{\langle\frac{L_1-L_{-1}}{2i} \rangle_{\Psi_\alpha(t)}}= h_\alpha \sinh(\rho_0)\sin(t)+O(c^0),
\eea
where we have used (\ref{L0global}). Our proposal of the radial momentum operator is 
\bea\label{radialmomentumoperator}
\hmp=\frac{L_1-L_{-1}+\bar L_1-\bar L_{-1}}{2i},
\eea 
which gives the expected relation 
\bea\label{expradialp}
\langle \hmp \rangle_{\Psi_\alpha(t)} =\Delta_\alpha \sinh(\rho_0)\sin(t)\simeq m \sinh(\rho_0)\sin(t).
\eea
Actually, we can take $\hmp$ as the generator of the radial transformation $g(\rho,0)$ since $g(\rho,0)=e^{-i\hmp \rho}$. 

We can construct the position operator $\hat \rho_r$ from the classical Hamiltonian of the partcile. By using (\ref{radialcanonical}) and (\ref{globalenergy}) the Hamiltonian with $\dot\phi=0$ is
\bea\label{HamiltonianPrho}
H=\cosh(\rho) \sqrt{m^2+P_\rho^2}.
\eea
Taking the Hamiltonian operator $\hat H=L_0+\bar L_0$ and radial momentum operator $\hmp$  (\ref{radialmomentumoperator}) into the above equation, one could obtain the operator $\hat \rho_r$ by solving the operator equation. This suggests the position operator $\hat \rho$ can be constructed as
\bea\label{radialposition}
\hat \rho:= \arccosh\left[ \hat A\right], \ \ \hat A:= \hat H (\Delta_\alpha^2+\hmp^2)^{-1/2}.
\eea
In the above expression  $\arccosh(\hat {A})$ is defined as $\sum_{n=1}^{\infty}a_n\hat A^{n}$ where $a_n$ are Taylor coefficients of the function $\arccosh (x)$. To make   $\arccosh(\hat {A})$  to be a well defined bounded operator the series expansion should be convergent in the sense of operator algebra. In this section we only focus on the expectation value of the operators in the state $\psiradialright$	. For our purpose  we would  take the operator $\arccosh(\hat {A})$ to be  a well defined operator if the expansion  $\sum_{n=0}^{\infty}a_n\langle\hat A^{n}\rangle_{\Psi_\alpha(t)}$ is finite. 
\subsection{Check of our proposal}

The operator $\hat A$ are polynomials in $\hat H$ and $\hmp$, which are associated with the energy momentum operator $T$ and $\bar T$. Roughly, the relation is $\hat H,\hmp \sim \int f T+\int \bar f \bar T$, where $f$ and $\bar f$ are some functions. The state $\psiradialright$ is explained as a moving bulk particle state, which obviously should be a geometric state. Therefore, using the factorization property  for the operators $\hat H$ and $\hmp$, we obtain
 \bea\label{Afactor}
 \langle \hat A\rangle_{\Psi_\alpha(t)}= \langle \hat H\rangle_{\Psi_\alpha(t)} (\Delta_\alpha^2+\langle \hmp\rangle_{\Psi_\alpha(t)}^2)^{-1/2}+O(c^{-1}).
 \eea
Similarly, the operator $\hat A$ also satisfies the factorization property 
\bea\label{factor}
\langle \hat A^n \rangle_{\Psi_\alpha(t)} =\langle \hat A \rangle_{\Psi_\alpha(t)}^n+O(c^{-1}).
\eea
 One could show this by direct calculations for a given $n$.
 
Taking (\ref{exphamradial}) and (\ref{expradialp}) into (\ref{Afactor}),  we have 
\bea\label{Aexpectation}
\langle \hat A \rangle_{\Psi_\alpha(t)}=\frac{1}{\sqrt{1-\tanh^2(\rho_0)\cos^2(t)}}+O(c^{-1}).
\eea
Using the above result and (\ref{factor}) we have the expected relation
\bea\label{radialrhoexpectation}
&&\langle \hat \rho_r \rangle_{\Psi_\alpha(t)} = \arccosh\left[ \frac{1}{\sqrt{1-\tanh^2(\rho_0)\cos^2(t)}}  \right]+O(c^{-1})\nonumber \\
&&\phantom{\langle \hat \rho_r \rangle_{\Psi_\alpha(t)}}=\arctanh[\tanh(\rho_0)\cos(t)]+O(c^{-1}).
\eea
 The expectation values of the operator $\hat \rho_r$ and  $\hat P_\rho$ in the state $\psiradialright$ give the classical results (\ref{radialsolution}) and (\ref{canonicalmomentumglobal}) at the leading order of $c$. 
 
 It is convenient to introduce the scaled momentum operator $\hat p_\rho:= \hmp/c$. They can be taken as the operators related to the particle with mass $m/c$. $\hat p_\rho$ are classical operators, since their expectation values  in the state $\psiradialright$ are finite in the limit $c\to \infty$.  We can also define more general operators $\hat X(\hat \rho_r ,\hat p_\rho)$. Consider two  arbitrary classical operators $\hat {X}(\hat \rho_r,\hat p_\rho)$ and $\hat Y(\hat \rho_r,\hat p_\rho)$, which are functions of $\hat \rho_r$ and $\hat p_\rho$.  We also have the factorization property
\bea\label{genralfactor}
&&\langle \hat X(\hat \rho_r,\hat p_\rho) \hat Y(\hat \rho_r,\hat p_\rho) \rangle_{\Psi_\alpha(t)}\nonumber \\
&&= X( \rho(t),p_\rho(t)) Y(\rho(t),p_\rho(t))+O(c^{-1}),
\eea
at the leading order of $c$, where $\rho(t),p_\rho(t):=P_\rho(t)/c$ are given by (\ref{basicrequirement}). The proof is similar as (\ref{factor}). Therefore, the classical operators behave as $c$-number in the state $\psiradialright$.  The  Newton constant $G$ or $1/c$ plays the role as the parameter $\hbar$ in quantum mechanics. 

The commutator $[\hat X, \hat Y]$ would also have a correspondence 	to the Poisson bracket $\{X,Y\}$.  
For the radial moving particle the phase space is 2-dimensional, for which $\rho$ and $p_\rho$ are canonical variables.  The classical Poisson brackets of two functions $X(\rho,p_\rho)$ and $Y(\rho,p_\rho)$ are defined as
\bea\label{poissonbracket}
\{X,Y\}:= \frac{\partial X}{\partial\rho}\frac{\partial Y}{\partial p_\rho}-\frac{\partial X}{\partial p_\rho}\frac{\partial Y}{\partial\rho}.
\eea
 One special case is the fundamental Poisson bracket  $\{\rho, p_\rho \}=1$.  
 Since we have constructed the position and momentum operators, their commutators can be evaluated by the Virasoro algebra. Our task is to show how to obtain the classical Poisson brackets from the quantum commutators. This is similar as the process that the quantum commutators reduce to classical Poisson brackets in the limit $\hbar \to  0$. 
 
 Let's begin with the fundamental bracket  $\{ \rho, P_\rho\}=1$. To evaluate the corresponding quantum commutator $[\hat \rho_r,\hmp]$, we need $[\hat A^n, \hmp]$. For $n=1$
\bea
[\hat A, \hmp ]=[\hat H,\hmp](m^2+P_\rho^2)^{-1/2}=-i\hmq (\Delta_\alpha^2+\hmp^2)^{-1/2},
\eea 
with 
\bea
\hmq:= \frac{L_1+L_{-1}+\bar L_1+\bar L_{-1}}{2}.
\eea
 For general $n$ it is not easy to write down the results.  However, if we consider the commutators in the state $\psiradialright$ the expression would be very simple at the leading order of $c$. By using the factorization property of operators, we have
\bea
&&\langle [\hat A^n,\hmp] \rangle_{\Psi_\alpha(t)}\nonumber \\
&& =n \langle[\hat H,\hmp]\hat H^{n-1}(\Delta_\alpha^2+\hmp^2)^{-n/2}\rangle_{\Psi_\alpha(t)} +O(c^{-1}) \nonumber \\
&&=-i n\langle \hat A^{n} \rangle_{\Psi_\alpha(t)} \langle \hmq \hat H^{-1} \rangle_{\Psi_\alpha(t)}+O(c^{-1}) \nonumber \\
&& =-i n \tanh(\rho)  \cosh^n(\rho)+O(c^{-1}) .
\eea 
In the above calculation we only keep the leading order $c$ results. In the last step we use (\ref{radialrhoexpectation}) and  the fact
\bea
&&\langle \hmq \hat H^{-1} \rangle_{\Psi_\alpha(t)}
=\frac{\sinh(\rho_0)\cos(t)}{\cosh(\rho_0)}+O(c^{-1})=\tanh(\rho)+O(c^{-1}).
\eea
Finally, we have the result
\bea
&&\langle [\hat \rho_r, \hmp ] \rangle_{\Psi_\alpha(t)}\nonumber \\
&&=\sum_{n} a_n\langle [\hat A^n, \hmp ] \rangle_{\Psi_\alpha(t)} \nonumber \\
&&=-i\tanh(\rho)\sum_{n} n a_n \cosh^n(\rho) +O(c^{-1}) \nonumber \\
&&=-i \tanh(\rho) \cosh(\rho) \frac{d }{d \cosh(\rho)}\sum_n  a_n \cosh^{n}(\rho)+O(c^{-1})\nonumber \\
&&=-i+O(c^{-1}).
\eea
The quantum commutator reduces to classical Poisson brackets as
\bea\label{quantumtoclassical}
\lim_{c\to \infty} ic \langle [\hat \rho_r, \hat p_\rho] \rangle_{\Psi_\alpha(t)} =\{\rho, p_\rho \}=1.
\eea 
The above result is consistent with the factorization property. The expectation values of $\hat \rho_r$ and $\hat p_\rho$ are of order $O(c^0)$. According to the factorization property (\ref{genralfactor}) the commutator $\langle [\hat \rho_r, \hat p_\rho] \rangle_{\Psi_\alpha(t)}$ should be vanishing at the order $O(c^0)$. The result (\ref{quantumtoclassical}) shows that the commutator is of $O(1/c)$.	
With the fundamental brackets one could derive the general Poisson brackets (\ref{poissonbracket}). 
\section{Locally excited state in CFT}\label{locallysection}
According to the extrapolate dictionary of AdS/CFT, the bulk operator $\phi_\alpha(\rho,\vec{x})$ and the dual boundary CFT operator $\mo(\vec{x})$ are related by
\bea
\mo (\vec{x})=\lim_{\rho \to \infty} e^{\rho \Delta_\alpha }
\phi_\alpha(\rho,\vec{x}), 
\eea
in the global coordinate.
We expect the bulk  state $\phi_\alpha (\rho,\vec{x})|0\rangle$ should reduce to the locally excited state $\mo(\vec{x})|0\rangle$ in CFT near the AdS boundary. To regularize this state $\mo(\vec{x})$ one could introduce a cut-off $\epsilon$ and define 
\bea\label{localCFTstate}
|\psi(\vec{x})\rangle_\epsilon := \mathcal{N}e^{-\epsilon \hat H} \mo (\vec{x})|0)\rangle,
\eea
where $\mathcal{N}$ is the normalization constant. 
We can take $\epsilon$ as the UV cut-off of theory with $\epsilon \ll 1$. The locally excited states has been studied in many literatures on the dynamics behavior of entanglement entropy, see \cite{Nozaki:2014hna}-\cite{Kudler-Flam:2020xqu} and references therein. 
In the following we would like to focus on such state, which is expected to be described by a point particle with the initial location near the boundary of AdS \cite{Berenstein:2019tcs}. 

Denote the boundary coordinate as $x^{\pm} = \phi\pm t$. By a Wick-rotation we have the Euclidean coordinate $w:=\phi-i \tau$ and $\bar w:= \phi+i\tau$. With a conformal mapping $z=e^{iw}$, the cylinder is mapped to $z$-plane. The state will be defined on the $z$-plane. The local state (\ref{localCFTstate}) inserted at $w_0=\bar w_0=0$ is given by
\bea\label{localradial}
|\psi_\alpha(z_0,\bar z_0)\rangle_\epsilon=\mathcal{N}(z_0,\bar z_0)\mo(z_0,\bar z_0)|0\rangle,
\eea
where $z_0=e^{i w_0}=e^{\epsilon},\bar z=e^{-i \bar w_0}=e^{\epsilon}$, and  normalization constant $\mathcal{N}(z_0,\bar z_0)=(z_0 z_0^*-1)^{h_\alpha} (\bar z_0 \bar z_0^*-1)^{\bar h_\alpha}$. We can also write the above state as 
\bea
|\psi_\alpha(z_0,\bar z_0)\rangle_\epsilon=\mathcal{N}(z_0,\bar z_0)e^{z_0L_{-1}+\bar z_0 \bar L_{-1}}|\mathcal{\mathcal{O}_\alpha}\rangle,
\eea
with  the primary state  $|\mathcal{O}_\alpha\rangle:=  \lim_{z\to 0}\mathcal{O}_\alpha |0\rangle$. 
For $h_\alpha \sim O(c)$, the energy of this state is
\bea\label{energylocalradial}
\psilocalleft \hat H \psilocalright= \frac{h_\alpha (z_0 z_0^*+1)}{z_0 z_0^*-1}+\frac{\bar h_\alpha (\bar z_0 \bar z_0^*+1)}{\bar z_0 \bar z_0^*-1},
\eea
where $\hat H=L_0+\bar L_0$. If $\epsilon \ll 1$ we have
\bea\label{energylocal}
\psilocalleft \hat H \psilocalright=\frac{\Delta_\alpha}{\epsilon}.
\eea
For the static particle located at the AdS boundary $\rho_0 \gg 1$, the energy is given by  (\ref{energyglobal}) with $v_\phi=0$, that is \bea
E= m\cosh (\rho_0)\simeq \frac{1}{2} \Delta_\alpha e^{\rho_0}. \eea
Comparing with (\ref{energylocal}) we obtain the relation  $
\frac{1}{\epsilon}\simeq \frac{1}{2}e^{\rho_0}$, which provides an interpretation of $\log (\frac{1}{\epsilon})$ as the initial location of the bulk particle. This is also consistent with the UV/IR relation in the context of AdS/CFT\cite{Susskind:1998dq}.

One could check the expectation value of momentum operator $P_\phi=L_0-\bar L_0$ in the state (\ref{localradial}) is vanishing. Therefore, we can interpret this state is dual to a particle moving in the radial direction in the bulk, that is $v_\phi=0$. \\

\subsection{State with angular momentum}
It is more interesting to construct the state with non-vanishing $v_\phi$. From (\ref{canonicalmomentumglobal}) and (\ref{globalenergy}) with $\rho_0 \gg 1$ we have
\bea\label{localarbitraryenergymomentum}
&&P_\phi\simeq \frac{m e^{\rho_0} v_\phi}{2\sqrt{1-v_\phi^2}},\nonumber \\
&&E\simeq \frac{me^{\rho_0}}{2\sqrt{1-v_\phi^2}}.
\eea
This motivates us to construct the state with non-vanishing $v_\phi$ by a boost with velocity $v_\phi$.  The coordinates transform as 
\bea
x'^{+}=e^{-\lambda} x^{+}, \quad x'^{-}=e^{\lambda} x^{-}.
\eea
where $\lambda$ is the rapidity with $ v_\phi=\tanh \lambda$.

We propose that the dual state of a moving bulk particle with initial position $(\rho,\phi)=(\rho_0, 0)$ ($\rho_0\gg 1$) and velocity $v_\phi$ is given by the locally excited state $\mathcal{O}_\alpha (w_0 e^{-\lambda},\bar w_0 e^{\lambda})|0\rangle$. On the $z$-plane the state is defined as
\bea\label{boostlocalstate}
|\psi_\alpha (z_\lambda,\bar z_{\lambda})\rangle_\epsilon:= \mathcal{N}(z_\lambda,\bar z_\lambda)e^{z_\lambda L_{-1}+\bar z_\lambda \bar L_{-1}}|\mathcal{\mathcal{O}_\alpha}\rangle,
\eea
where $z_\lambda=e^{i w_0e^{-\lambda}}=e^{\epsilon e^{-\lambda}},\bar z_\lambda=e^{-i \bar w_0e^{\lambda}}=e^{\epsilon e^{\lambda}}$, and  normalization constant $\mathcal{N}(z_\lambda,\bar z_\lambda)=(z_\lambda z_\lambda^*-1)^{h_\alpha} (\bar z_\lambda \bar z_\lambda^*-1)^{\bar h_\alpha}$.

The energy can be obtained by the replacement  $z_0\to z_\lambda$, $\bar z_0 \to \bar z_\lambda$ in (\ref{energylocalradial}). Keeping the leading order of $\epsilon$ we have
\bea
E=~_\epsilon \langle \psi_\alpha (z_\lambda,\bar z_{\lambda} )| (L_0+\bar L_0)|\psi_\alpha (z_\lambda,\bar z_{\lambda})\rangle_\epsilon=\frac{\Delta_\alpha}{\epsilon}\cosh \lambda.
\eea 
Similarly, the angular momentum is given by
\bea
P_\phi=~_\epsilon \langle \psi_\alpha (z_\lambda,\bar z_{\lambda} )| (L_0-\bar L_0)|\psi_\alpha (z_\lambda,\bar z_{\lambda})\rangle_\epsilon=\frac{\Delta_\alpha}{\epsilon}\sinh \lambda. 
\eea
These are consistent with the results (\ref{localarbitraryenergymomentum}).

Consider the time evolution  and define the time-dependent state
\bea
|\psi_\alpha(t)\rangle_\epsilon :=U_g(t)|\psi_\alpha (z_\lambda,\bar z_{\lambda})\rangle_\epsilon,
\eea  	
where $U_g=e^{i H_g t}$. It is obvious that the energy and angular momentum  are independent with $t$. 
\subsection{Angular coordinate operator}
We can construct the position and momentum operator as we have done for the radial moving case. We will show they can be expressed as operator functions of the global Virasoro generators.

To simplify the notation  the expectation value of operator $\hat X$ in the state $|\psi_\alpha(t)\rangle_\epsilon$ is denoted by $\langle \hat X\rangle_{\psi_\alpha(t)}$.  The following formulas are useful for our construction,
\bea\label{usefulformula}
&&\langle L_1 \rangle_{\psi_\alpha(t)}=\frac{2 h_\alpha z_\lambda e^{it}}{z_\lambda z_\lambda^*-1}\simeq \frac{h_\alpha }{\epsilon}e^{\lambda+it},\nonumber \\
&&\langle \bar L_1 \rangle_{\psi_\alpha(t)}=\frac{2 \bar h_\alpha \bar z_\lambda e^{it}}{\bar z_\lambda \bar z_\lambda^*-1}\simeq \frac{\bar h_\alpha }{\epsilon}e^{-\lambda+it},\nonumber \\
&&\langle L_{-1} \rangle_{\psi_\alpha(t)}=\frac{2 h_\alpha z_\lambda e^{-it}}{z_\lambda z_\lambda^*-1}\simeq \frac{h_\alpha }{\epsilon}e^{\lambda-it},\nonumber \\
&&\langle \bar L_{-1} \rangle_{\psi_\alpha(t)}=\frac{2 \bar h_\alpha \bar z_\lambda e^{-it}}{\bar z_\lambda \bar z_\lambda^*-1}\simeq \frac{\bar h_\alpha }{\epsilon}e^{-\lambda-it}.
\eea
In previous section we have defined two Hermitian operators $\hat \mathcal{P}_\rho$ and $\hat \mathcal{Q}_\rho$ which are linear combinations of the global Virasoro generators. Let's define two more independent operators
\bea
&&\hat \mathcal{S}_\phi:= \frac{L_{1}-L_{-1}-(\bar L_{1}-\bar L_{-1})}{2i},\nonumber \\
&&\hat \mathcal{T}_\phi:= \frac{L_{1}+L_{-1}-(\bar L_{1}+\bar L_{-1})}{2}.
\eea
By using (\ref{usefulformula}) it is straightforward to evaluate the expectation values of the four Hermitian operators. The results are
\bea\label{PQSTexpectation}
&&\langle \hat \mathcal{P}_\rho\rangle_{\psi_\alpha(t)}=\frac{\Delta_\alpha}{\epsilon}\cosh \lambda \sin t, \nonumber \\
&&\langle \hat \mathcal{Q}_\rho\rangle_{\psi_\alpha(t)}=\frac{\Delta_\alpha}{\epsilon}\cosh \lambda \cos t, \nonumber \\
&&\langle \hat \mathcal{S}_\phi\rangle_{\psi_\alpha(t)}=\frac{\Delta_\alpha}{\epsilon}\sinh \lambda \sin t, \nonumber \\
&&\langle \hat \mathcal{T}_\phi \rangle_{\psi_\alpha(t)}=\frac{\Delta_\alpha}{\epsilon}\sinh \lambda \cos t.
\eea 
The above results are consistent with the radial moving case $\lambda=0$.\\
Since the state (\ref{boostlocalstate}) is a geometric state, the global Virasoro generators also satisfy the factorization property. 

Now we move to construct the operator $\hat \phi$. The expectation $\langle \hat \phi \rangle_{\psi_\alpha(t)}$ is expected to give the classical solution (\ref{radialsolution}). Rewrite  (\ref{radialsolution}) as
\bea
\phi =\arccos\left[ \frac{1}{\sqrt{1+(\tanh \lambda \tan t)^2}}\right] =\arccos \left[\frac{\cosh \lambda \cos t}{\sqrt{(\cosh \lambda \cos t)^2+ (\sinh \lambda \sin t)^2}}\right].
\eea
The angular coordinate operator $\hat \phi$ is suggested to be
\bea\label{localarbitraryphi}
\hat \phi =\arccos \hat B ,\quad \hat B=\hat \mathcal{Q}_\rho (\hat \mathcal{Q}_\rho^2 +\hat \mathcal{S}_\phi^2)^{-1/2},
\eea
where $\arccos \hat B := \sum_n b_n \hat B^n$, $b_n$ are Taylor coefficients of the function $\arccos(x)$. Using (\ref{PQSTexpectation}) we have the result
\bea\label{localB}
\langle \hat B\rangle_{\psi_\alpha(t)}=\frac{1}{\sqrt{1+(\tanh \lambda \tan t)^2}}+O(c^{-1}).
\eea

One could check that $\langle \hat \phi \rangle_{\psi_\alpha(t)}=\phi(t)$. For the radial moving particle we have 
$\langle \hat \phi \rangle_{\Psi_\alpha(t)} =0$.
\subsection{Radial momentum and coordinate operator}
For $\rho_0\gg 1$ the radial momentum is 
\bea\label{radiallocalclassical}
P_\rho(t)\simeq \frac{\Delta_\alpha}{\epsilon}\frac{\sqrt{1-v_{\phi }^2} \sin (t) }{\sqrt{1+v_{\phi }^2\tan ^2(t) } },
\eea
which is different from $\langle \hat \mathcal{P}_\rho\rangle_{\psi_\alpha(t)}$ (\ref{PQSTexpectation}). We should include more terms to produce the above expected result. By using (\ref{PQSTexpectation}) we  suggest the following radial momentum operator
\bea\label{localradialmomentum}
\hat P_\rho:= \hat \mathcal{P}_\rho \cos \hat \phi-\hat \mathcal{T}_\rho \sin \hat \phi.
\eea
By the definition of $\hat \phi$ we have $\sin \hat \phi =\sqrt{1-\cos^2 \hat \phi}=\sqrt{1-\hat B^2}$, which can be written as
\bea
\sin\hat \phi=\hat \mathcal{S}_\phi (\hat \mathcal{Q}_\rho^2+\hat \mathcal{S}_\phi^2)^{-1/2}.
\eea 
The expectation value of $\hat P_\rho$ is given by 
\bea
\langle \hat P_\rho \rangle_{\psi_\alpha(t)}=\frac{\Delta_\alpha}{\epsilon}\frac{\sin t}{\cosh \lambda \sqrt{1+\tanh^2\lambda \tan^2 t}}+O(c^0,\epsilon^0),
\eea
which is equal to (\ref{radiallocalclassical}). 
For the special radial moving case $\langle \hat \phi \rangle_{\Psi_\alpha(t)}=0$, we can effectively take $\hat \mathcal{P}_\rho$ as the radial momentum operator.

The radial position operator $\hat \rho$ can be constructed by using the relation
\bea
H=\coth \rho\sqrt{P_\phi^2+(m^2+P_\rho^2)\sinh^2\rho}.
\eea
 The solution of  the above equation for $\rho$ actually gives an ansatz of the radial position operator $\hat \rho$. 
We have constructed the Hamiltonian operator $\hat H$, the angular momentum operator $\hat P_\phi$ and the radial momentum operator $\hat P_\rho$.
Taking them into the solution we obtain 
\bea\label{localoperatorrho}
&&\hat \rho=\frac{1}{2}\arccosh (\hat r), \nonumber \\
&& \hat r:=\Big\{(\hat H+\hat P_\phi)(\hat H-\hat P_\phi)+[(\hat H+\hat P_\phi)^2-\Delta_\alpha^2-\hat P_\rho^2]^{1/2}\nonumber \\
&&\phantom{\hat r:=\Big\{} [(\hat H-\hat P_\phi)^2-\Delta_\alpha^2-\hat P_\rho^2]^{1/2}\Big\} (\Delta_\alpha^2+\hat P_\rho^2)^{-1}
\eea
One could check the above expression will become (\ref{radialposition}) for $\hat P_\phi=0$. With some  calculations we can find the expected relation 
\bea
\langle \hat \rho \rangle_{\psi_\alpha(t)}=\arctanh[ \sqrt{(1-v_\phi^2)\cos^2 t+v_\phi^2}]+O(c^{-1},\epsilon).
\eea
\subsection{Poisson brackets}

As a check of our proposals  we will show how to get the Poisson brackets from the position and momentum operators. The phase space of the bulk moving particle is 4-dimensional with the canonical variables $\{\phi,\rho,P_\phi,P_\phi\}$. We will focus on the fundamental Poisson brackets. 

Firstly, consider the commutator $[\hat \phi,\hat P_\phi]$. With the definitions we have the following commutation relations,
\bea\label{QScom}
&&[\hat \mathcal{Q}_\rho, \hat P_\phi]=i \hat \mathcal{S}_\phi, \quad
[\hat \mathcal{S}_\phi, \hat P_\phi]=-i \hat \mathcal{Q}_\rho,\nonumber \\
&&[\hat \mathcal{P}_\rho, \hat P_\phi]=-i \hat \mathcal{T}_\phi, \quad
[\hat \mathcal{T}_\phi, \hat P_\phi]=i \hat \mathcal{P}_\rho,\nonumber \\
&&[\hat \mathcal{P}_\rho, \hat \mathcal{Q}_\phi]=-i \hat H, \quad
[\hat \mathcal{P}_\rho, \hat \mathcal{S}_\phi]=0,\nonumber \\
&&[\hat \mathcal{T}_\phi, \hat \mathcal{S}_\phi]=i \hat H, \quad
[\hat \mathcal{T}_\phi, \hat \mathcal{Q}_\phi]=0, \nonumber \\
&&[\hat \mathcal{P}_\rho, \hat H]=-i \hat \mathcal{Q}_\rho, \quad
[\hat \mathcal{T}_\phi, \hat H]=i \hat \mathcal{S}_\phi,\nonumber \\
&&[\hat \mathcal{Q}_\rho, \hat H]=i \hat \mathcal{P}_\rho, \quad
[\hat \mathcal{S}_\phi, \hat H]=-i \hat \mathcal{T}_\phi.
\eea
With these and the definition of $\hat B$ (\ref{localarbitraryphi}) we can obtain 
\bea
&&\langle [\hat B ,\hat P_\phi ]\rangle_{\psi_\alpha(t)}\nonumber \\
&&=\langle \Big\{ [\hat \mathcal{Q}_\rho, \hat P_\phi](\hat \mathcal{Q}_\rho^2 +\hat \mathcal{S}_\phi^2)^{-1/2}+ \hat \mathcal{Q}_\rho[(\hat \mathcal{Q}_\rho^2 +\hat \mathcal{S}_\phi^2)^{-1/2},\hat P_\phi]\Big\}\rangle_{\psi_\alpha(t)}\nonumber \\
&&=i\langle  \hat \mathcal{S}_\phi(\hat \mathcal{Q}_\rho^2 +\hat \mathcal{S}_\phi^2)^{-1/2}\rangle_{\psi_\alpha(t)},
\eea
where in the second step we have used 
\bea
&&\langle [(\hat \mathcal{Q}_\rho^2 +\hat \mathcal{S}_\phi^2)^{-1/2},\hat P_\phi] \rangle_{\psi_\alpha(t)}\nonumber \\
&& =-\langle (\hat \mathcal{Q}_\rho^2 +\hat \mathcal{S}_\phi^2)\rangle_{\psi_\alpha(t)}^{-1}\langle [(\hat \mathcal{Q}_\rho^2 +\hat \mathcal{S}_\phi^2)^{1/2},\hat P_\phi] \rangle_{\psi_\alpha(t)}+O(c^{-1})\nonumber \\
&&=-\frac{1}{2} \langle (\hat \mathcal{Q}_\rho^2 +\hat \mathcal{S}_\phi^2)\rangle_{\psi_\alpha(t)}^{-3/2}\langle [\hat \mathcal{Q}_\rho^2 +\hat \mathcal{S}_\phi^2,\hat P_\phi] \rangle_{\psi_\alpha(t)}+O(c^{-1})\nonumber \\
&&=-\frac{1}{2} \langle (\hat \mathcal{Q}_\rho^2 +\hat \mathcal{S}_\phi^2)\rangle_{\psi_\alpha(t)}^{-3/2}\langle \left(\hat \mathcal{Q}_\rho [\hat \mathcal{Q}_\rho,\hat P_\phi]+ [\hat \mathcal{Q}_\rho,\hat P_\phi]\hat \mathcal{Q}_\rho+\hat \mathcal{S}_\rho [\hat \mathcal{S}_\rho,\hat P_\phi]+ [\hat \mathcal{S}_\rho,\hat P_\phi]\hat \mathcal{S}_\rho \right)\rangle_{\psi_\alpha(t)}+O(c^{-1})\nonumber \\
&&=0+O(c^{-1}),
\eea
where the last step follows from  (\ref{QScom}).  In the above evaluation we have used the factorization property. Therefore, the equality is established only in the leading order of $c$. \\
Now we can evaluate the commutator
\bea
&&\langle [\hat \phi,\hat P_\phi] \rangle_{\psi_\alpha(t)}=\sum_n b_n \langle [\hat B^n,\hat P_\phi] \rangle_{\psi_\alpha(t)}\nonumber \\
&&\phantom{\langle [\hat \phi,\hat P_\phi] \rangle_{\psi_\alpha(t)}}=\sum_n n b_n\langle \hat B\rangle_{\psi_\alpha(t)}^{n-1} \langle [\hat B,\hat P_\phi]\rangle_{\psi_\alpha(t)}+O(c^{-1})\nonumber \\
&&\phantom{\langle [\hat \phi,\hat P_\phi] \rangle_{\psi_\alpha(t)}}=i\frac{\partial }{\partial \langle \hat B\rangle_{\psi_\alpha(t)}} \arccos (\langle \hat B\rangle_{\psi_\alpha(t)})\langle  \hat \mathcal{S}_\phi(\hat \mathcal{Q}_\rho^2 +\hat \mathcal{S}_\phi^2)^{-1/2}\rangle_{\psi_\alpha(t)}+O(c^{-1})\nonumber \\
&&\phantom{\langle [\hat \phi,\hat P_\phi] \rangle_{\psi_\alpha(t)}}=-i\frac{1}{\sqrt{1-(\langle \hat B\rangle_{\psi_\alpha(t)})^2}}\langle  \hat \mathcal{S}_\phi(\hat \mathcal{Q}_\rho^2 +\hat \mathcal{S}_\phi^2)^{-1/2}\rangle_{\psi_\alpha(t)}+O(c^{-1})\nonumber \\
&&\phantom{\langle [\hat \phi,\hat P_\phi] \rangle_{\psi_\alpha(t)}}=-i+O(c^{-1}).
\eea
In the last step we have used (\ref{PQSTexpectation}) and (\ref{localB}).

Let's introduce the scaled operator $\hat p_\phi:=\hat P_\phi/c$. The quantum commutator reduces to Poisson brackets as
\bea
\lim_{c\to \infty}i c\langle [\hat \phi,\hat p_\phi] \rangle_{\psi_\alpha(t)}=1.
\eea
Using the above result we can evaluate the more general commutators such as  
\bea\label{fphipcom}
\lim_{c\to \infty } ic\langle [F(\hat \phi),\hat p_\phi] \rangle_{\psi_\alpha(t)}=\langle \frac{\partial F(\hat \phi)}{\partial \hat \phi} \rangle_{\psi_\alpha(t)},
\eea
where $F(\hat \phi)$ is arbitrary functions of $\hat \phi$. One could derive the above expression by taking $\hat p_\phi$ as $\frac{i}{c}\frac{\partial }{\partial \hat \phi}$ when evaluating the commutators.

Now let's consider the commutator $[\hat p_\phi,\hat p_\rho]$, where we 
define the scaled radial momentum operator $\hat p_\rho:= \hat P_\rho/c$. Using the definition of $\hat P_\rho$ (\ref{localradialmomentum}) and commutation relations (\ref{QScom} ) and (\ref{fphipcom}), we have
\bea
&&\lim_{c\to \infty}i c\langle [\hat p_\rho,\hat p_\phi] \rangle_{\psi_\alpha(t)}\nonumber \\
&&=\lim_{c\to \infty} i  \langle [ \hat \mathcal{P}_\rho \cos \hat \phi-\hat \mathcal{T}_\rho \sin \hat \phi,\hat p_\phi] \rangle_{\psi_\alpha(t)}\nonumber \\
&&=\lim_{c\to \infty} i\langle \Big\{ [ \hat \mathcal{P}_\rho, \hat p_\phi]\cos \hat \phi+\hat \mathcal{P}_\rho [\cos \hat \phi,\hat p_\phi] -[\hat \mathcal{T}_\rho ,\hat p_\phi]\sin \hat \phi -\hat \mathcal{T}_\rho [\sin \hat \phi,\hat p_\phi]\Big\} \rangle_{\psi_\alpha(t)}\nonumber \\
&&=0.
\eea
With some calculations one can also get the following commutators,
\bea\label{calculationinappendix}
&&\lim_{c\to \infty}i c\langle [\hat p_\rho,\hat \phi] \rangle_{\psi_\alpha(t)}=0,\nonumber \\
&&\lim_{c\to \infty}i c\langle [\hat \rho,\hat p_\phi] \rangle_{\psi_\alpha(t)}=0,\nonumber\\
&&\lim_{c\to \infty}i c\langle [\hat \rho,\hat \phi] \rangle_{\psi_\alpha(t)}=0,\nonumber \\
&&\lim_{c\to \infty}i c\langle [\hat \rho,\hat p_\rho] \rangle_{\psi_\alpha(t)}=1.
\eea
We show the details of the calculation in Appendix.\ref{appcom}
\subsection{Equation of motion}\label{EoM}

With the Poisson brackets one could easily obtain the equation of motion of the particle. 
For a classical system with canonical variable $\{q_i,p_i\}$ the Hamiltonian equation can be expressed as 
\bea
\frac{d q_i}{dt}=\{ q_i,H\},\quad \frac{d p_i}{dt}=\{ p_i,H\},
\eea
where $H$ is the Hamiltonian of the system. In the previous section we have constructed the position and momentum operators and shown the quantum commutators can reduce to the classical Poisson brackets in the limit $c\to \infty$. Now we would like to find the classical equation in the same limit. \\
Take $\hat \phi$ as an example. Let's define the scaled Hamiltonian operator $\hat h:= \hat H/c$. Using the result (\ref{bhcom}) and $\hat \phi:= \arccos \hat B$ we have
\bea
&&\lim_{c\to \infty} ic \langle [\phi,\hat h]\rangle_{\psi_\alpha(t)}\nonumber \\
&&=-\lim_{c\to \infty} i (1-\langle \hat B\rangle_{\psi_\alpha(t)}^2)^{-1/2}\langle [\hat B,\hat H]\rangle_{\psi_\alpha(t)}\nonumber \\
&&=\lim_{c\to \infty}\langle (\hat \mathcal{P}_\rho \hat \mathcal{S}_\phi+\hat \mathcal{Q}_\rho\hat \mathcal{T}_\phi)(\hat \mathcal{Q}_\rho^2 +\hat \mathcal{S}_\phi^2)^{-1} \rangle_{\psi_\alpha(t)}\nonumber \\
&&=\frac{\sinh \lambda  \cosh \lambda }{(\sinh \lambda  \sin t)^2+(\cosh \lambda  \cos t)^2},
\eea
where in the last step we have used (\ref{PQSTexpectation}). The final result is same as $\frac{d\langle \hat \phi \rangle_{\psi_\alpha(t)}}{dt}$. Therefore, we find the equation of motion
\bea
\frac{d\langle \hat \phi \rangle_{\psi_\alpha(t)}}{dt}=\lim_{c\to \infty} ic \langle [\phi,\hat h]\rangle_{\psi_\alpha(t)},
\eea
which can be taken as the classical limit of the Heisenberg equation. 

Another example is the operator $\hat P_\rho$. Using (\ref{prhoHcom}) and (\ref{PQSTexpectation}) one could check the equation of motion
\bea
\frac{d\langle \hat p_\rho\rangle_{\psi_\alpha(t)}}{dt}=\lim_{c\to \infty} ic \langle [\hat p_\rho,\hat h]\rangle_{\psi_\alpha(t)}.
\eea 
The interested reader can check the other two equations associated with $\hat \rho$ and $\hat P_\phi$.

\section{Conclusion and discussion}

The main result of our paper is to explore the CFT dual of a bulk moving particle. As mentioned in the introduction the problem has two aspects. 

Firstly, we construct the state that are expected to be dual to the moving particle. Two examples are shown. For the radial moving particle starting from arbitrary position $\rho_0$ we find the CFT state can be described by the regularized bulk local states that are discussed in previous paper \cite{Goto:2017olq}. The other one is the boundary locally excited state. This state can be explained as particle starting from  the AdS boundary.   In this case we find  the dual state of the particle with angular momentum can be related to a boost. The rapidity of the boost is associated with the velocity in the $\phi$ direction. As far as we know the state with a boost hasn't been discussed in other papers. 

The other aspect of the problem is to construct the position and momentum operators associated with the particle. We should also note that the operators in the radial moving example are special case of $\{\rho,\hat P_\rho,\hat \phi, \hat P_\phi\}$. However, we haven't successfully constructed the state dual to arbitrary moving particles in the bulk. It would be interesting to find such states and check whether the constructed operators could give the correct results. 

Generally, we have no systematical method to find the operators. Therefore, we should discuss case by case at present. Of course, there are some basic constraints on the constructions.

 Let us summarize some important clues.
  The energy and angular momentum momentum of the particle should be related to the Hamiltonian and momentum operators of the dual CFTs. The particle can be taken as excited state of the bulk. Hence, the energy and angular momentum of the particle should be equal to the difference between the excited state and the background state. In the CFTs they should be related to the expectation value of the Hamiltonian and momentum operators. 
  
The basic requirement is that the expectation value of the constructed operators in the dual CFT states should give the classical solution at the leading order of $G$. Actually, this is the important guidance to the constructions. For the examples that are shown in our paper are simple, since we could find the exact classical solution. We find the operators can be constructed only by using the stress energy tensor, in fact only by the global Virasoso generators. The reason is that the dual state can be obtained by global symmetry. For the background state beyond the vacuum the symmetry is broken. We don't expect these operators should be universe, that is independent with the background geometry. But the stress energy tensor should be the building block of the position and momentum operators. 

Another requirement is the quantum commutators of the constructed operators should reduce to the classical Poisson brackets in the semiclassical limit $c\to \infty$. This can also be seen as a check of our constructions. In both examples in our paper we show the correspondence between the quantum and classical brackets. 

There are many important and interesting problems that we haven't touch in this paper. In the following we will briefly discuss three such problems that are worthy to study in the future.

\subsection{Other examples}
We only focus on the vacuum AdS$_3$ in the global coordinate. It is easily to generalize to other situations, such as the Poincare coordinate, AdS Rindler. One can use the similar methods that we have used. It is an interesting excise to work out the results in different coordinate and compare with our results. In particular, the AdS Rindler will help us to understand more on the entanglement wedge construction or subregiona/subregion duality\cite{Wall:2012uf}-\cite{Dong:2016eik}. 
For example, the Poincare coordinate of pure AdS$_3$ is 
\bea
ds^2=\frac{dy^2-dt_p^2+dx^2}{y^2}.
\eea
The  state $|\Psi_\alpha\rangle $ in this case  corresponds to the bulk local excitation at point $(t_p,y,x)=(0,1,0)$.  The bulk local state at point $(t_p,y,x)=(0,y,x)$ is given by 
\bea
|\Psi_\alpha(y,x)\rangle= g(y,x)|\Psi_\alpha\rangle,
\eea
where
\bea
g(y,x):= e^{i\left( L_0+\frac{L_1+L_{-1}}{2}\right)x^{+}}e^{-i\left( \bar L_0+\frac{\bar L_1+\bar L_{-1}}{2}\right)x^{-}} y^{\frac{L_1-L_{-1}+\bar L_1-\bar L_{-1}}{2}},
\eea
where $x^{\pm}:=x\pm t_p$.
The time evolution is controlled by the operator $U_p:= e^{i(L_{-1}^h+\bar L^h_{-1})t_p}$, where $L^h_{-1}:=-L_0+\frac{L_1+L_{-1}}{2}$ and $\bar L_{-1}^h:=-\bar L_0+\frac{\bar L_1+\bar L_{-1}}{2}$ are the generators in the hyperbolic basis as shown in \cite{Goto:2017olq}. By similar method we expect one could obtain the position and momentum operators in the Poincar\'e coordinate.  It would be interesting to see the difference and relation with the global coordinate.

More interesting case is BTZ black hole as the background geometry. One could use the thermofield double states to work out the results. In this case there is a horizon in the bulk. It is interesting to explore how to construct the corresponding operators once the particle is inside the horizon. 

The generalization to higher dimension is not so straightforward, but we expect one could have a similar construction as the 3D AdS. In the vacuum case the operators are only associated with the global symmetry. In higher dimension the dual CFT also has global conformal symmetry. But in higher dimension the situation will be more complicated, thus more interesting phenomena are expected to appear.
\subsection{Coordinate dependence}

 In the last section we discuss the generalization to other coordinate such as the Poincar\'e AdS. 
 The vacuums in different coordinate are not same. The particle states are actually differnt in these coordinate.
 
It is obvious that our constructions of position and momentum operators depend on the coordinate.  Even in the global coordinate, one could choose different coordinates. The operators should depend on the canonical variables that one choose. 

For example, in the radial moving case one could choose $D:=\arccosh \rho$ as the position coordinate of the particle.
The Lagrangian is 
\bea
L(t)=-m\sqrt{D^2-\frac{\dot{D}^2}{D^2-1}-(D^2-1)\dot{\phi}^2}.
\eea 
Hence, the canonical momentum is given by
\bea
P_D:=\frac{\partial L}{\partial{\dot{D}}}=\frac{m \dot{D}}{(D^2-1)\sqrt{D^2-\frac{\dot{D}^2}{D^2-1}-(D^2-1)\dot{\phi}^2}}.
\eea
The Hamiltonian of the classical particle with $\dot \phi=0$ is given by
\bea\label{hamD}
H= D\sqrt{m^2+(D^2-1)P_D}.
\eea
Thanks to the factorization property of the geometric sate,  we could guess the position operator should be 
\bea\label{hamopeD}
\hat D:= \cosh \hat \rho =\hat A =\hat H (\Delta_\alpha^2 +\hmp^2)^{-1/2}.
\eea
The equation (\ref{hamD}) gives the expression of the canonical momentum operator
\bea
\hat P_D= \hmp (\Delta_\alpha^2+\hmp^2)^{1/2}(\hat H^2-\Delta_\alpha^2-\hmp^2)^{-1/2}.
\eea 
One could check the above operators by comparing the expectation value of them with the classical geodesic solution.

The coordinate transformation is a special case of the canonical transformation of the phase space of the classical particle. One could choose any set of the canonical variables. And the corresponding position and momentum operators can be constructed by the same methods as above.

\subsection{How to understand gravity in the CFTs?}

Our results provide a framework to explore the explanation of gravity in the CFTs.  In general relativity the dynamics of the particle is given by the geodesic equation. The geodesic equation is equal to equation of motion for the system with Lagrangian $L= -m\int d\tau $, where $\tau$ is the proper time of the particle.

In our approach we assume the existence of the CFT operators that are dual to the  canonical variables of the particle. Moreover, we construct the operators,  the expectation values of which give the particle's position and momentum.  A remarkable fact is that all these operators are constructed by the Hermitian operators $\hmp,\hmq,\hms,\hmt$, $\hat H$ and $\hat P_\phi$ , which are independent linear combinations of global Virasoro generators.

We also show how to obtain the classical Poisson brackets from the quantum commutators of the constructed operators. The Hamiltonian equations, i.e., the geodesic equations, can be expressed by the Poisson brackets as we have shown in section.\ref{EoM}. Hence, the dynamics of the bulk particle is determined by the commutation relations algebra (\ref{QScom}).  Further, these relations can be derived from the stress energy tensor commutators, i.e., $[T_{\mu\nu},T_{\rho\sigma}]$.  The geodesic equation should have a correspondence to the stress energy tensor commutator, at least in our special cases.  However, we have no evidence to conclude that the correspondence is also true for general background geometry. It is worth to study more general examples and make the correspondence more explicit. 

Recently, Susskind proposes the size-momentum correspondence, that gives a connection between radial momentum of a bulk particle, operator size and complexity\cite{Susskind:2018tei}. The size of the operator is expected to be proportional to the radial momentum of a bulk moving particle. One could refer to \cite{Susskind:2018tei}-\cite{Susskind:2019ddc} for the definition of these concepts. The operator size can be evaluated in SYK models\cite{Roberts:2018mnp}\cite{Qi:2018bje}. For general theories, such as holographic field theory,  it is expected the operator size may be  associated with a Hermitian operator\cite{Mousatov:2019xmc}\cite{Magan:2020iac}. Follow the notation of \cite{Mousatov:2019xmc}, the size of operator $\mathcal{O}$ that act on reference state $|\Psi\rangle$ is given by 
\bea
S_{|\Psi\rangle}(\mathcal{O}):=\frac{\langle\Psi|\mathcal{O}^\dagger \hat S_{|\Psi\rangle} \mathcal{O}|\Psi\rangle}{\langle\Psi|\mathcal{O}^\dagger \mathcal{O}|\Psi\rangle},
\eea
where $\hat S_{|\Psi\rangle}$ is expected to be a semi-definite, Hermitian operator. 

Actually, in our approach the radial momentum of the bulk moving particle is also given by the expectation value of the operator $\hat P_\rho$. It is very interesting to check whether the operator size operator $\hat S_{|\Psi\rangle}$ has some connection with the radial momentum operator $\hat P_\rho$ in our paper. \\
~\\

~\\
~\\
{\bf Acknowledgements} I would like to thank Qin Qin for useful discussions.  I am supposed by the National Natural
Science Foundation of China under Grant No.12005070 and the Fundamental Research
Funds for the Central Universities under Grants NO.2020kfyXJJS041
~\\
~\\
\appendix
\section{Commutators}\label{appcom}
In this section we will show the details of the calculation of commutators(\ref{calculationinappendix}). \\
Consider the commutator $\langle [\hat p_\rho,\hat \phi]\rangle_{\psi_\alpha(t)}$. By definition we have
\bea\label{app1}
&&\langle[\hat P_\rho,\hat \phi]\rangle_{\psi_\alpha(t)}\nonumber \\
&&= \langle[\hat \mathcal{P}_\rho \cos \hat \phi-\hat \mathcal{T}_\rho \sin \hat \phi,\hat \phi]\rangle_{\psi_\alpha(t)}\nonumber \\
&&= \langle[\hat \mathcal{P}_\rho ,\hat \phi]\cos \hat \phi-[\hat \mathcal{T}_\rho,\hat \phi] \sin \hat \phi\rangle_{\psi_\alpha(t)}.\nonumber 
\eea
It is useful to evaluate $\langle[\hat \mathcal{P}_\rho,\hat B]\rangle_{\psi_\alpha(t)}$. By using (\ref{QScom}) we have
\bea
&&\langle[\hat \mathcal{P}_\rho,\hat B]\rangle_{\psi_\alpha(t)}\nonumber \\
&&=\langle[\hat \mathcal{P}_\rho,\hat \mathcal{Q}_\rho (\hat \mathcal{Q}_\rho^2 +\hat \mathcal{S}_\phi^2)^{-1/2}]\rangle_{\psi_\alpha(t)}\nonumber \\
&&=\langle[\hat \mathcal{P}_\rho,\hat \mathcal{Q}_\rho](\hat \mathcal{Q}_\rho^2 +\hat \mathcal{S}_\phi^2)^{-1/2}\rangle_{\psi_\alpha(t)} +\langle\hat \mathcal{Q}_\rho[\hat \mathcal{P}_\rho,(\hat \mathcal{Q}_\rho^2 +\hat \mathcal{S}_\phi^2)^{-1/2}]\rangle_{\psi_\alpha(t)} \nonumber\\
&&=-i\langle\hat H(\hat \mathcal{Q}_\rho^2 +\hat \mathcal{S}_\phi^2)^{-1/2}\rangle_{\psi_\alpha(t)} -\frac{1}{2}\langle\hat \mathcal{Q}_\rho[\hat \mathcal{P}_\rho,(\hat \mathcal{Q}_\rho^2 +\hat \mathcal{S}_\phi^2)](\hat \mathcal{Q}_\rho^2 +\hat \mathcal{S}_\phi^2)^{-3/2}\rangle_{\psi_\alpha(t)} +O(c^{-1})\nonumber\\
&&=-i\langle \hat H \hat \mathcal{S}_\phi^2 (\hat \mathcal{Q}_\rho^2 +\hat \mathcal{S}_\phi^2)^{-3/2}\rangle_{\psi_\alpha(t)} +O(c^{-1}).
\eea
Similarly, we have
\bea
&&\langle[\hat \mathcal{T}_\rho,\hat B]\rangle_{\psi_\alpha(t)}\nonumber \\
&&=\langle[\hat \mathcal{T}_\rho,\hat \mathcal{Q}_\rho (\hat \mathcal{Q}_\rho^2 +\hat \mathcal{S}_\phi^2)^{-1/2}]\rangle_{\psi_\alpha(t)}\nonumber \\
&&=\langle[\hat \mathcal{T}_\rho,\hat \mathcal{Q}_\rho](\hat \mathcal{Q}_\rho^2 +\hat \mathcal{S}_\phi^2)^{-1/2}\rangle_{\psi_\alpha(t)} +\langle\hat \mathcal{Q}_\rho[\hat P_\rho,(\hat \mathcal{Q}_\rho^2 +\hat \mathcal{S}_\phi^2)^{-1/2}]\rangle_{\psi_\alpha(t)} \nonumber\\
&&= -\frac{1}{2}\langle\hat \mathcal{Q}_\rho[\hat \mathcal{T}_\rho,(\hat \mathcal{Q}_\rho^2 +\hat \mathcal{S}_\phi^2)](\hat \mathcal{Q}_\rho^2 +\hat \mathcal{S}_\phi^2)^{-3/2}\rangle_{\psi_\alpha(t)} +O(c^{-1})\nonumber\\
&&=-i\langle \hat H \hat \mathcal{S}_\phi \mathcal{Q}_\rho(\hat \mathcal{Q}_\rho^2 +\hat \mathcal{S}_\phi^2)^{-3/2}\rangle_{\psi_\alpha(t)} +O(c^{-1}).
\eea
Using the above results we can obtain (\ref{app1}), the result is
\bea
&&\langle[\hat P_\rho,\hat \phi]\rangle_{\psi_\alpha(t)}\nonumber \\
&&=\sum_n b_n \langle \Big\{ \langle[\hat \mathcal{P}_\rho ,\hat B^n]\cos \hat \phi-[\hat \mathcal{T}_\rho,\hat B^n] \sin \hat \phi\Big\}\rangle_{\psi_\alpha(t)}\nonumber \\
&&=\sum_n n b_n \langle\Big\{ \langle[\hat \mathcal{P}_\rho ,\hat B]\hat B^{n-1}\cos \hat \phi-[\hat \mathcal{T}_\rho,\hat B]\hat B^{n-1} \sin \hat \phi\Big\}\rangle_{\psi_\alpha(t)}+O(c^{-1})\nonumber \\
&&=\sum_n n b_n \langle\hat B^{n-1} \hat H \hat \mathcal{S}_\phi (\hat \mathcal{Q}_\rho^2 +\hat \mathcal{S}_\phi^2)^{-3/2}\Big\{ -i\hat \mathcal{S}_\phi\cos \hat \phi+i\hat \mathcal{Q}_\rho \sin \hat \phi\Big\}\rangle_{\psi_\alpha(t)}+O(c^{-1})\nonumber \\
&&=0+O(c^{-1}),
\eea
where in the last step we use the fact $\langle \hat \mathcal{S}_\phi\cos \hat \phi-\hat \mathcal{Q}_\rho \sin \hat \phi\rangle_{\psi_\alpha(t)}=O(c^{-1})$.\\
Therefore, we have the classical Poisson bracket 
\bea
\lim_{c\to\infty} i c\langle[\hat p_\rho,\hat \phi]\rangle_{\psi_\alpha(t)}=0.
\eea
Now let's turn to the commutator $[\hat \rho,\hat p_\phi]$. We should calculate $[\hat r, \hat p_\phi]$, where $\hat r$ is defined as (\ref{localoperatorrho}). With some tedious but straightforward calculations we find $[\hat r, \hat p_\phi]\propto \langle [\hat p_\rho,\hat p_\phi]\rangle_{\psi_\alpha(t)}$. We have shown that $\lim_{c\to \infty} ic \langle [\hat p_\rho,\hat p_\phi]\rangle_{\psi_\alpha(t)}=0$. One could conclude that 
\bea
\lim_{c\to \infty} ic \langle [\hat \rho,\hat p_\phi]\rangle_{\psi_\alpha(t)}=0.
\eea
We leave it as an exercise for interested reader to prove $\lim_{c\to \infty}i c\langle [\hat \rho,\hat \phi] \rangle_{\psi_\alpha(t)}=0$. \\
Finally, let's show the last commutator in (\ref{calculationinappendix}). $\hat \rho$ is a function of $\hat H$,$\hat P_\phi$ and $\hat P_\rho$. We will need the commutator $[\hat P_\rho,\hat H]$, which can be  written as
\bea\label{PHcomrelation1}
&&[\hat P_\rho,\hat H]=[ \hat \mathcal{P}_\rho \cos \hat \phi-\hat \mathcal{T}_\rho \sin \hat \phi,\hat H]\nonumber \\
&&\phantom{[\hat P_\rho,\hat H]}=[ \hat \mathcal{P}_\rho \hat B-\hat \mathcal{T}_\rho (1-\hat B^2)^{1/2},\hat H]\nonumber \\
&&\phantom{[\hat P_\rho,\hat H]}=[ \hat \mathcal{P}_\rho ,\hat H]\hat B+\hat \mathcal{P}_\rho[  \hat B,\hat H]-[ \hat \mathcal{T}_\phi ,\hat H](1-\hat B^2)^{1/2}-\hat \mathcal{T}_\phi[(1-\hat B^2)^{1/2},\hat H]\nonumber \\
&&\phantom{[\hat P_\rho,\hat H]}=-i\hat \mathcal{Q}_\rho\hat B+\hat \mathcal{P}_\rho[  \hat B,\hat H]-i\mathcal{S}_\phi(1-\hat B^2)^{1/2}-\hat \mathcal{T}_\phi[(1-\hat B^2)^{1/2},\hat H]\nonumber \\
\eea
The expectation value of $[\hat P_\rho,\hat H]$ is
\bea\label{prhoHcom}
&&\langle [\hat P_\rho,\hat H]\rangle_{\psi_\alpha(t)}\nonumber \\
&&=-i \langle \left(\hat \mathcal{Q}_\rho^4+\hat \mathcal{Q}_\rho^2(2\hat \mathcal{S}_\phi^2-\hat \mathcal{T}_\phi^2 )-2\hat \mathcal{Q}_\rho  \hat \mathcal{P}_\rho\hat \mathcal{S}_\phi\hat \mathcal{T}_\phi  +\hms^2(\hms^2-\hmp^2)\right)\left(\hmq^2+\hms^2 \right)^{-3/2}\rangle_{\psi_\alpha(t)}+O(c^0),\nonumber \\
\eea
where we have used 
\bea\label{bhcom}
&&\langle [\hat B,\hat H]\rangle_{\psi_\alpha(t)}\nonumber \\
&&=i\langle (\hat \mathcal{P}_\rho \hat \mathcal{S}_\phi+\hat \mathcal{Q}_\rho\hat \mathcal{T}_\phi)\hat \mathcal{S}_\phi (\hat \mathcal{Q}_\rho^2 +\hat \mathcal{S}_\phi^2)^{-3/2} \rangle_{\psi_\alpha(t)}+O(c^{-1}).
\eea
Now recall the definition of $\hat r$ (\ref{localoperatorrho}),  we obtain the commutator
\bea\label{rprhocom}
&&\langle [\hat r,\hat P_\rho]\rangle_{\psi_\alpha(t)}\nonumber \\
&&=-2 \langle \Big\{\hat H [\hat P_\rho,\hat H]\left( \hat H^2-\hat P_\phi^2-\Delta_\alpha^2-\hat P_\rho^2 \right)(\hat P_\rho^2+\Delta_\alpha^2)^{-1}\nonumber \\
&&\phantom{=}\left( (\hat H^2-\hat P_\phi^2-\Delta_\alpha^2-\hat P_\rho^2)^2-4 \hat P_\phi^2 (\hat P_\rho^2+\Delta_\alpha^2) \right)^{-1/2}\Big\}\rangle_{\psi_\alpha(t)}+O(c^{-1}).\nonumber \\
\eea
Thus we have
\bea
\langle [\hat \rho,\hat P_\rho] \rangle_{\psi_\alpha(t)}=\frac{1}{\sqrt{\langle \hat r\rangle_{\psi_\alpha(t)}^2-1}}\langle[\hat r,\hat P_\rho] \rangle_{\psi_\alpha(t)}+O(c^{-1})
\eea
Taking (\ref{rprhocom}) into the above expression and replacing the operator by their expectation, we finally obtain the expected result
\bea
\langle [\hat \rho,\hat P_\rho] \rangle_{\psi_\alpha(t)}=-i+O(c^{-1}).
\eea
Therefore, the classical Poisson bracket is given by
\bea
\lim_{c\to \infty}i c\langle [\hat \rho,\hat P_\rho] \rangle_{\psi_\alpha(t)}=1.
\eea

\end{document}